\newif\ifarxiv

\arxivtrue

\documentclass[conference, a4paper]{IEEEtran}
\IEEEoverridecommandlockouts
\usepackage{cite}
\usepackage{acro}

\DeclareAcronym{URLLC}{
  short = URLLC ,
  long  = Ultra-Reliable Low Latency Communication ,
  class = abbrev
}

\DeclareAcronym{SINR}{
    short = {SINR} ,
    long  = Signal to Interference plus Noise Ratio ,
    class = abbrev
}

\DeclareAcronym{LR}{
    short = {LR} ,
    long  = Logistic Regression ,
    class = abbrev
}

\DeclareAcronym{MAP}{
    short = {MAP} ,
    long  = Maximum A Posteriori ,
    class = abbrev
}

\DeclareAcronym{AWGN}{
    short = {AWGN} ,
    long  = Additive White Gaussian Noise ,
    class = abbrev
}

\DeclareAcronym{BI-AWGN}{
    short = {BI-AWGN} ,
    long  = Binary-Input Additive White Gaussian Noise ,
    class = abbrev
}

\DeclareAcronym{LOS}{
    short = {LOS} ,
    long  = Line-Of-Sight ,
    class = abbrev
}

\DeclareAcronym{NLOS}{
    short = {NLOS} ,
    long  = Non-Line-Of-Sight ,
    class = abbrev
}

\DeclareAcronym{TDL-D}{
    short = {TDL-D} ,
    long  = Tapped Delay Line D ,
    class = abbrev
}

\DeclareAcronym{TDL}{
    short = {TDL} ,
    long  = Tapped Delay Line ,
    class = abbrev
}

\DeclareAcronym{CDL}{
    short = {CDL} ,
    long  = Clustered Delay Line ,
    class = abbrev
}

\DeclareAcronym{CDL-D}{
    short = {CDL-D} ,
    long  = Clustered Delay Line D ,
    class = abbrev
}

\DeclareAcronym{MDS}{
    short = {MDS} ,
    long  = Maximum Distance Separable ,
    class = abbrev
}

\DeclareAcronym{LR-LLR}{
    short = {LR-LLR} ,
    long  = Logistic Regression on LogLikelihood Ratios ,
    class = abbrev
}

\DeclareAcronym{AUC}{
    short = {AUC} ,
    long  = Area Under the false-positive-Curve ,
    class = abbrev
}

\DeclareAcronym{SLSQP}{
    short = {SLSQP} ,
    long  = Sequential Least Squares Programming ,
    class = abbrev
}

\DeclareAcronym{TH-SNR}{
    short = {LR-SNR} ,
    long  = Logistic Regression on Signal-to-Noise-Ratio ,
    class = abbrev
}

\DeclareAcronym{Q-SNR}{
    short = {Q-SNR} ,
    long  = Quantized Signal-to-Noise-Ratio ,
    class = abbrev
}

\DeclareAcronym{LR-SC}{
    short = {LR-SC} ,
    long  = Logistic Regression on Subcode features ,
    class = abbrev
}

\DeclareAcronym{TH-LLR}{
    short = {Q-LLR} ,
    long  = Quantized LogLikelihood Ratio ,
    class = abbrev
}

\DeclareAcronym{DIDA}{
    short = {DA2SGMM} ,
    long  = Dual Autoencoding 2-Stage Gaussian Mixture Model ,
    class = abbrev
}

\DeclareAcronym{IR}{
    short = {IR} ,
    long  = Incremental Redundancy ,
    class = abbrev
}

\DeclareAcronym{EEG}{
    short = {EEG} ,
    long  = Energy Efficiency Gain ,
    class = abbrev
}

\DeclareAcronym{DMRS}{
  short = DMRS ,
  long  = Demodulation Reference Signal,
  class = abbrev
}

\DeclareAcronym{WI}{
  short = WI ,
  long  = Work Item,
  class = abbrev
}

\DeclareAcronym{QAM}{
  short = QAM ,
  long  = Quadrature Amplitude Modulation ,
  class = abbrev
}

\DeclareAcronym{OFDM}{
  short = OFDM ,
  long  = Orthogonal Frequency Division Multiplexing ,
  class = abbrev
}

\DeclareAcronym{OFDMA}{
  short = OFDMA ,
  long  = Orthogonal Frequency Division Multiplexing Access ,
  class = abbrev
}

\DeclareAcronym{E2E}{
  short = E2E ,
  long  = End-to-End ,
  class = abbrev
}

\DeclareAcronym{DL}{
  short = DL ,
  long  = DownLink ,
  class = abbrev
}

\DeclareAcronym{PDCP}{
  short = PDCP ,
  long  = Packet Data Convergence Protocol ,
  class = abbrev
}

\DeclareAcronym{prHARQ}{
  short = prHARQ ,
  long  = proactive HARQ with prediction ,
  class = abbrev
}

\DeclareAcronym{eprHARQ}{
  short = eprHARQ ,
  long  = early proactive HARQ with prediction ,
  class = abbrev
}

\DeclareAcronym{paHARQ}{
  short = paHARQ ,
  long  = proactive HARQ ,
  class = abbrev
}

\DeclareAcronym{reHARQ}{
  short = reHARQ ,
  long  = reactive HARQ ,
  class = abbrev
}

\DeclareAcronym{GF}{
  short = GF ,
  long  = Grant-Free ,
  class = abbrev
}

\DeclareAcronym{VLF}{
  short = VLF ,
  long  = Variable-Length Feedback ,
  class = abbrev
}

\DeclareAcronym{SED}{
  short = SED ,
  long  = Stong Error Detection ,
  class = abbrev
}

\DeclareAcronym{ML}{
  short = ML ,
  long  = Maximum Likelihood ,
  class = abbrev
}

\DeclareAcronym{RB}{
  short = RB ,
  long  = Resource Block ,
  class = abbrev
}

\DeclareAcronym{RAN}{
  short = RAN ,
  long  = Radio Access Network ,
  class = abbrev
}

\DeclareAcronym{CSI}{
  short = CSI ,
  long  = Channel State Information ,
  class = abbrev
}

\DeclareAcronym{CSIT}{
  short = CSIT ,
  long  = Channel State Information at the Transmitter ,
  class = abbrev
}

\DeclareAcronym{CSIR}{
  short = CSIR ,
  long  = Channel State Information at the Receiver ,
  class = abbrev
}

\DeclareAcronym{MCS}{
  short = MCS ,
  long  = Modulation and Coding Scheme ,
  class = abbrev
}

\DeclareAcronym{CFI}{
  short = CFI ,
  long  = Control Format Indicator ,
  class = abbrev
}

\DeclareAcronym{UL}{
  short = UL ,
  long  = UpLink ,
  class = abbrev
}

\DeclareAcronym{SR}{
  short = SR ,
  long  = Scheduling Request ,
  class = abbrev
}

\DeclareAcronym{BER}{
  short = BER ,
  long  = Bit Error Rate ,
  class = abbrev
}

\DeclareAcronym{BLER}{
  short = BLER ,
  long  = Block Error Rate ,
  class = abbrev
}

\DeclareAcronym{CRC}{
  short = CRC ,
  long  = Cyclic Redundancy Check ,
  class = abbrev
}

\DeclareAcronym{RRH}{
  short = RRH ,
  long  = Remote Radio Head ,
  class = abbrev
}

\DeclareAcronym{BBU}{
  short = BBU ,
  long  = BaseBand Unit ,
  class = abbrev
}

\DeclareAcronym{CB}{
  short = CB ,
  long  = Code Block ,
  class = abbrev
}

\DeclareAcronym{CBG}{
  short = CBG ,
  long  = Code Block Group ,
  class = abbrev
}

\DeclareAcronym{R-CBG}{
  short = R-CBG ,
  long  = Reduced Code Block Group ,
  class = abbrev
}

\DeclareAcronym{AR-CBG}{
  short = AR-CBG ,
  long  = Adaptive Reduced Code Block Group ,
  class = abbrev
}

\DeclareAcronym{TB}{
  short = TB ,
  long  = Transport Block ,
  class = abbrev
}

\DeclareAcronym{BG2}{
  short = BG2 ,
  long  = Base Graph 2 ,
  class = abbrev
}

\DeclareAcronym{BG1}{
  short = BG1 ,
  long  = Base Graph 1 ,
  class = abbrev
}

\DeclareAcronym{SNR}{
  short = SNR ,
  long  = Signal-to-Noise Ratio ,
  class = abbrev
}

\DeclareAcronym{CCE}{
  short = CCE ,
  long  = Control Channel Element ,
  class = abbrev
}

\DeclareAcronym{PDCCH}{
  short = PDCCH ,
  long  = Physical Downlink Control Channel ,
  class = abbrev
}

\DeclareAcronym{PUCCH}{
  short = PUCCH ,
  long  = Physical Uplink Control Channel ,
  class = abbrev
}

\DeclareAcronym{LTE}{
  short = LTE ,
  long  = Long Term Evolution ,
  class = abbrev
}

\DeclareAcronym{NGMN}{
  short = NGMN ,
  long  = Next Generation Mobile Networks ,
  class = abbrev
}

\DeclareAcronym{RNTI}{
  short = RNTI ,
  long  = Radio Network Temporary Identifier ,
  class = abbrev
}

\DeclareAcronym{FC}{
  short = FC ,
  long  = Fully Connected ,
  class = abbrev
}

\DeclareAcronym{3GPP}{
  short = 3GPP ,
  long  = 3rd Generation Partnership Project ,
  class = abbrev
}

\DeclareAcronym{HRLLC}{
  short = HRLLC ,
  long  = High-Reliable Low Latency Communication ,
  class = abbrev
}

\DeclareAcronym{SIMO}{
  short = SIMO ,
  long  = single-input multiple-output ,
  class = abbrev
}

\DeclareAcronym{MIMO}{
  short = MIMO ,
  long  = Multiple-Input Multiple-Output ,
  class = abbrev
}

\DeclareAcronym{TTI}{
  short = TTI ,
  long  = Transmission Time Interval ,
  class = abbrev
}

\DeclareAcronym{sTTI}{
  short = sTTI ,
  long  = short Transmission Time Interval ,
  class = abbrev
}

\DeclareAcronym{RTT}{
  short = RTT ,
  long  = Round Trip Time ,
  class = abbrev
}

\DeclareAcronym{LDPC}{
  short = LDPC ,
  long  = Low-Density Parity-Check ,
  class = abbrev
}

\DeclareAcronym{UE}{
  short = UE ,
  long  = User Equipment ,
  class = abbrev
}

\DeclareAcronym{TI}{
    short = TI ,
    long  = Tactile Internet ,
    class = abbrev
}

\DeclareAcronym{BS}{
  short = BS ,
  long  = Base Station ,
  class = abbrev
}

\DeclareAcronym{FPR}{
  short = FPR ,
  long  = False-Positive Rate ,
  class = abbrev
}

\DeclareAcronym{FNR}{
  short = FNR ,
  long  = False-Negative Rate ,
  class = abbrev
}

\DeclareAcronym{DCI}{
  short = DCI ,
  long  = Downlink Control Information ,
  class = abbrev
}

\DeclareAcronym{HARQ}{
  short = HARQ ,
  long  = Hybrid Automatic Repeat reQuest ,
  class = abbrev
}

\DeclareAcronym{IIOT}{
  short = IIOT ,
  long  = Industrial Internet of Things ,
  class = abbrev
}

\DeclareAcronym{RV}{
  short = RV ,
  long  = Redundancy Version ,
  class = abbrev
}

\DeclareAcronym{ACK}{
  short = ACK ,
  long  = ACKnowledgment ,
  class = abbrev
}

\DeclareAcronym{NACK}{
  short = NACK ,
  long  = Non-ACKnowledgment ,
  class = abbrev
}

\DeclareAcronym{CG}{
  short = CG ,
  long  = Configured Grant ,
  class = abbrev
}

\DeclareAcronym{M-TRP}{
  short = M-TRP ,
  long  = Multi-TRP ,
  class = abbrev
}

\DeclareAcronym{E-HARQ}{
  short = E-HARQ ,
  long  = Early HARQ ,
  class = abbrev
}

\DeclareAcronym{P-HARQ}{
  short = P-HARQ ,
  long  = Predictive incremental-redundancy rateless HARQ ,
  class = abbrev
}

\DeclareAcronym{R-HARQ}{
  short = R-HARQ ,
  long  = Rateless incremental-redundancy HARQ ,
  class = abbrev
}

\DeclareAcronym{C-RAN}{
  short = C-RAN ,
  long  = Cloud Radio Access Network ,
  class = abbrev
}

\DeclareAcronym{C-HARQ}{
  short = C-HARQ ,
  long  = Conventional incremental-redundancy HARQ ,
  class = abbrev
}

\DeclareAcronym{mMTC}{
  short = mMTC ,
  long  = massive Machine Type Communications ,
  class = abbrev
}

\DeclareAcronym{5G}{
  short = 5G ,
  long  = Fifth Generation ,
  class = abbrev
}

\DeclareAcronym{SPS}{
  short = SPS ,
  long  = Semi-Persistent Scheduling ,
  class = abbrev
}

\DeclareAcronym{PI}{
  short = PI ,
  long  = Pre-emption Indication ,
  class = abbrev
}

\DeclareAcronym{V2X}{
  short = V2X ,
  long  = Vehicle-To-Everything ,
  class = abbrev
}

\DeclareAcronym{VR}{
  short = VR ,
  long  = Virtual Reality ,
  class = abbrev
}

\DeclareAcronym{NR}{
  short = NR ,
  long  = New Radio ,
  class = abbrev
}

\DeclareAcronym{eMBB}{
  short = eMBB ,
  long  = enhanced Mobile BroadBand ,
  class = abbrev
}

\DeclareAcronym{LLR}{
  short = LLR ,
  long  = Log-Likelihood Ratio ,
  class = abbrev
}

\DeclareAcronym{VNR}{
  short = VNR ,
  long  = Variable Node Reliability ,
  class = abbrev
}

\DeclareAcronym{BSC}{
  short = BSC ,
  long  = Binary Symmetric Channel ,
  class = abbrev
}

\DeclareAcronym{angelsperarea}{
  short = \ensuremath{a} ,
  long  = The number of angels per unit area ,
  sort  = a ,
  class = nomencl
}
\DeclareAcronym{numofangels}{
  short = \ensuremath{N} ,
  long  = The number of angels per needle point ,
  sort  = N ,
  class = nomencl
}
\DeclareAcronym{areaofneedle}{
  short = \ensuremath{A} ,
  long  = The area of the needle point ,
  sort  = A ,
  class = nomencl
}
\usepackage[cmex10]{amsmath}  
\usepackage{amssymb,amsfonts,amsthm}
\usepackage{algorithmic}
\usepackage{graphicx}
\usepackage{textcomp}
\usepackage{xcolor}
\usepackage{bbm}

\usepackage[utf8]{inputenc} 
\usepackage[T1]{fontenc}
\usepackage{url}              
\usepackage{cite}             

\usepackage{mleftright}       
\mleftright                   

\usepackage{booktabs}         

\def\BibTeX{{\rm B\kern-.05em{\sc i\kern-.025em b}\kern-.08em
    T\kern-.1667em\lower.7ex\hbox{E}\kern-.125emX}}

\DeclareMathOperator*{\argmax}{arg\,max}

\DeclareMathOperator*{\esssup}{ess\,sup}
\DeclareMathOperator*{\essinf}{ess\,inf}

\newtheorem{thrm}{Theorem}
\newtheorem{lemma}{Lemma}
\newtheorem{cor}{Corollary}
\newtheorem{prop}{Proposition}
\newtheorem{definition}{Definition}

\begin{document}

\title{On the Limits of HARQ Prediction for Short Deterministic Codes with Error Detection in Memoryless Channels \ifarxiv(Extended Version with Proofs)\fi
\thanks{The last author acknowledges the financial support by the Federal Ministry of Education and Research of Germany in the programme of “Souverän. Digital. Vernetzt.” Joint project 6G-RIC, project identification number: 16KISK020K.}}

\author{%
\IEEEauthorblockN{Bar{\i}{\c s} G{\" o}ktepe\IEEEauthorrefmark{1},
Cornelius Hellge\IEEEauthorrefmark{1},
Tatiana Rykova\IEEEauthorrefmark{1},
Thomas Schierl\IEEEauthorrefmark{1} and
Slawomir Stanczak\IEEEauthorrefmark{1}\IEEEauthorrefmark{2}}
\IEEEauthorblockA{\IEEEauthorrefmark{1}Fraunhofer Heinrich Hertz Institute,
Berlin, Germany\\
\IEEEauthorrefmark{2}Technische Universität Berlin,
Berlin, Germany\\
Email: first.last@hhi.fraunhofer.de}
}

\maketitle

\begin{abstract}
We provide a mathematical framework to analyze the limits of Hybrid Automatic Repeat reQuest (HARQ) and derive analytical expressions for the most powerful test for estimating the decodability under maximum-likelihood decoding and $t$-error decoding. Furthermore, we numerically approximate the most powerful test for sum-product decoding. We compare the performance of previously studied HARQ prediction schemes and show that none of the state-of-the-art HARQ prediction is most powerful to estimate the decodability of a partially received signal vector under maximum-likelihood decoding and sum-product decoding. Furthermore, we demonstrate that decoding in general is suboptimal for predicting the decodability.
\end{abstract}


\section{Introduction} 
\ac{HARQ} is a widely used physical layer retransmission mechanism to ensure high reliability, while not sacrificing too much spectral efficiency. Generally, \ac{HARQ} can be regarded as a special case of \ac{VLF} codes. In \cite{5961844}, Polyanskiy \emph{et al.} show that \ac{VLF} codes improve the achievable rate significantly. Particularly, stop-feedback codes attract particular interest, due to their simple stop feedback mechanism. For these codes, Polyanskiy \emph{et al.} provide a random-coding bound on the achievable performance, assuming a noiseless and zero-latency feedback channel. This analysis is refined by Östman \emph{et al.} in \cite{8989314} by considering the latter two constraints.\\
In particular, the feedback latency becomes an issue with regards to \ac{URLLC} use cases, where for 6G, end-to-end latencies of down to sub-milliseconds are foreseen \cite{9390169}. To address the latency issue, \ac{HARQ} prediction mechanisms have been studied in the literature \cite{caire_fast_harq, zhou_ieeetran, phdthesis_csi_harq, snr_prediction, prediction_cran1, prediction_cran2, deep_ml_eharq_iq, early_harq_schemes2,early_harq_schemes, prediction_nn, llr_channel_harq_prediction, ldpc_subcodes, journal_eharq_paper, feedback_prediction_iiot}. \ac{HARQ} prediction provides the feedback ahead of the end of the transmission, such that the feedback is available at the transmitter when it has to decide whether more redundancy is required or not. In particular, the literature can be divided into schemes that rely on \ac{CSI} for the decodability estimation and schemes that use the partially received signal vector. The first class of schemes is well understood also in theoretic terms. In \cite{caire_fast_harq}, Makki et. al. derive closed-form expressions for the message decoding probabilities and the performance in quasi-static Rician and Rayleigh fading channels. In \cite{prediction_cran1}, Khalili and Simeone conduct a theoretical analysis for \ac{HARQ} feedback prediction in \acp{C-RAN}. However, these schemes cannot be transfered to memoryless channels, where \ac{CSI} knowledge cannot be determined at the receiver. In contrast to that, \ac{LLR} and subcode-based schemes that use the partially received signal vector can be applied to memoryless channels. In fading channels, multiple works have shown a superior performance of these schemes \cite{feedback_prediction_iiot, early_harq_schemes2, arxiv_eharq_journal}. Nevertheless, the limits of these schemes are not well understood and no theoretical analysis exists to date.\\

\emph{Contributions:} In this work, we provide a mathematical framework to analyze the limits of \ac{HARQ} prediction. In contrast to previous theoretic works that try to design \ac{VLF} coding schemes, we consider the channel code and the associated decoder as a given. We derive the most powerful tests for arbitrary deterministic codes under \ac{ML} decoders, particularizing for the \ac{AWGN} channel, and $t$-error correcting codes under hard decision decoding. Furthermore, we numerically explore the performance of the most powerful test for a $(3,6)$-regular \ac{LDPC} code under \ac{ML} decoding, $t$-error hard decision decoding and sum-product decoding. We also compare our results to actual decoding of the subcodeword, also known as proactive \ac{HARQ} \cite{gf_harq_oulu, gf_harq_aalborg}, as a prediction strategy for a fixed $\alpha$-level. We obtain results that show that decoding may not be an optimal strategy to predict the decodability of a reception in the most cases. Finally, we compare the performance of other state-of-the-art decodability prediction schemes to the most powerful test and show that none of these achieves the attainable power.\\

\emph{Notation:} Throughout our work, scalars are denoted by lower case letters, e.g. $\alpha$, vectors by lower case and bold letters, e.g. $\mathbf{y}$, matrices by upper case and bold letters, e.g. $\mathbf{B}$ and sets by calligraphic font, e.g. $\mathcal{C}$. We use upper case letters, e.g. $Y$, to denote random variables. The dimension of their multivariate counterparts is denoted by using superscripts, such as $Y^p := [Y_1, ..., Y_p]$. The realizations of random variables are noted in lower case letters, e.g. $Y^p = \mathbf{y}$. $\mathcal{N}(\mu, \sigma^2)$ designates the normal distribution with mean $\mu$ and variance $\sigma^2$. Furthermore, $P_{Y^p}$ and $P_{Y^p|X^p=\mathbf{c}_i}$ designate the probability measure associated to $Y^p$ and its counterpart condtioned on $X^p=\mathbf{c}_i$, respectively. With some abuse of notation, we designate by $P_{Y^p|X^n=\mathbf{c}}$, the probability of $Y^p$ given only the first $p$ symbols of the codeword $\mathbf{c}$. Additionally, we assume that for any $\mathbf{c}_i$ $P_{Y^p|X^p=\mathbf{c}_i}$ is absolutely continuous to a probability measure $\mu$, e.g. the Lebesgue measure or the counting measure. Finally, $\mathbf{I}_k$ designates a unit matrix of size $k \times k$.

\section{System Model}
\begin{figure}
    \centering
    \includegraphics[width=0.8\columnwidth]{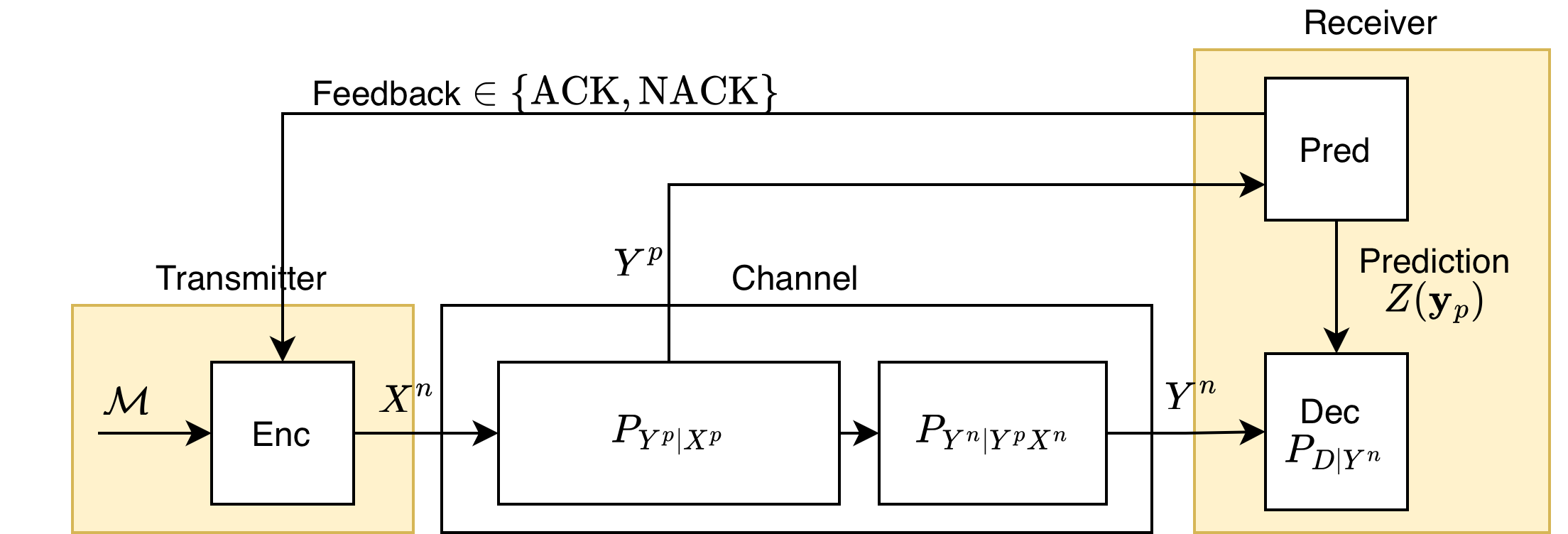}
    \caption{System model of \ac{HARQ} prediction. The encoder maps a valid message $\mathcal{M}$ to a codeword $X^n$, which is transmitted over a channel $P_{Y^n|X^n}$. Furthermore, the predictor taps the channel $Y^p$ after $p$ symbols to predict the success of the decoder $P_{D|Y^n}$, where the decoder $D$ chooses based on $Y^n$ a message from the extended set $\mathcal{M}_\mathrm{x}$.}
    \label{fig:model}
\end{figure}
Fig.~\ref{fig:model} shows the system model of \ac{HARQ} feedback prediction. Let us consider the channel input and output sets as $\mathcal{A}^n := \mathcal{A}^p \times \mathcal{A}^r$ and $\mathcal{B}^n := \mathcal{B}^p \times \mathcal{B}^r$, respectively, where $n$ is the codeword length, $p$ is the prediction length and $r := n - p$. Furthermore, we consider a conditional probability measure, characterizing a memoryless channel, $ P_{Y^n|X^n} := P_{Y^p|X^p} P_{Y^r| X^n} : \mathcal{A}^n \mapsto \mathcal{B}^p \times \mathcal{B}^r$, a set of message candidates $\mathcal{M}_\mathrm{x} = \{1,2,...,|\mathcal{M}_\mathrm{x}|\}$, an arbitrary codebook $\mathcal{C}_\mathrm{x} := \{\mathbf{c}_i \in \mathcal{A}^n : i \in \mathcal{M}_\mathrm{x}\}$ and a corresponding decoder $D \in \mathcal{M}_\mathrm{x}$ characterized by a conditional probability measure $P_{D|Y^n} : \mathcal{B}^n \mapsto \mathcal{M}_\mathrm{x}$. Furthermore, in practice error detection codes, such as \ac{CRC}, are commonly used. Essentially, these define a set $\mathcal{M} \subseteq \mathcal{M}_\mathrm{x}$ of messages that are considered to be valid. The output distribution $P_{Y^p}$ designates, unless stated otherwise, the distribution generated only by the valid codewords in $\mathcal{M}$. Also, we define the events
for successful and unsuccessful decoding of $D$ as $\mathcal{D}_\mathrm{A} \subset [0,1] \times \mathcal{M} \times \mathcal{B}^n$ and $\mathcal{D}_\mathrm{N}\subset [0,1] \times \mathcal{M} \times \mathcal{B}^n$, respectively, with $\mathcal{D}_\mathrm{A} \cap \mathcal{D}_\mathrm{N} = \emptyset$ and $\mathbb{P}[\mathcal{D}_\mathrm{A}] = \sum_{i\in\mathcal{M}}\mathbb{P}[X^n = \mathbf{c}_i]\mathbb{P}[D = i | X^n = \mathbf{c}_i] = 1 - \mathbb{P}[\mathcal{D}_\mathrm{N}]$. Then, the Neyman-Pearson fundamental lemma guarantees the existence of a most powerful test $Z \in \{0, 1\}$ to distinguish between undecodable ($Z= 0$) and decodable ($Z=1$) signal vectors attaining the following minimum
\begin{align}\label{eq:np}
    \beta_\alpha(P_{Y^p|\mathcal{D}_\mathrm{A}}, P_{Y^p|\mathcal{D}_\mathrm{N}}) = \min_{Z : \mathbb{E}(Z | \mathcal{D}_\mathrm{A}) \geq 1 - \alpha} \mathbb{E}(Z | \mathcal{D}_\mathrm{N})\,,
\end{align}
where we designate $\alpha$ as the significance level of the test and $\beta_\alpha$ as its power \cite{lehmann2005testing}. The Neyman-Pearson fundamental lemma further guarantees the existence of such a test for all $\alpha$-levels. However, depending on the probability measures, this test may be randomized.
In the following theorem, we provide a test that attains the minimum of (\ref{eq:np}), i.e. is most powerful, for all $\alpha$-levels.
\begin{thrm}\label{th:np}
There exists a constant $c \in [0, 1/P_{\mathcal{D}_\mathrm{A}}]$ and a random variable $Z_\tau$, which takes the value $1$ with probability $\tau$ and $0$ otherwise, such that the minimum in (\ref{eq:np}) is attained by
\begin{align}\label{eq:test}
    Z_\mathrm{NP}(\mathbf{y}_p) := \begin{cases}
                1&\text{if }T_\mathrm{NP}(\mathbf{y}_p) > c,\\
                Z_\tau&\text{if }T_\mathrm{NP}(\mathbf{y}_p) = c,\\
                0&\text{if }T_\mathrm{NP}(\mathbf{y}_p) < c,\\
            \end{cases}
\end{align}
for any arbitrary $\alpha \in [0,1]$ with $T_\mathrm{NP}(\mathbf{y}_p) := \frac{dP_{Y^p|\mathcal{D}_\mathrm{A}}}{dP_{Y^p}}(\mathbf{y}_p)$, where $dP_{Y^p}$ is the output distribution density of the code and $dP_{Y^p|\mathcal{D}_\mathrm{A}}$ given as
\begin{align}\label{eq:pda}
    dP_{Y^p|\mathcal{D}_\mathrm{A}} &= \sum_{i \in \mathcal{M}}\frac{ dP_{Y^p|X^n}(\mathbf{y}_p | \mathbf{c}_i) P_{D|Y^pX^n}(i|\mathbf{y}_p, \mathbf{c}_i)}{|\mathcal{M}| P_{D|X^n}(i|\mathbf{c}_i)}\,.
\end{align}
\end{thrm}
\ifarxiv
\begin{proof}
    See App.~\ref{app:th1}.
\end{proof}
\fi
The following derivations apply to continuous as well as discrete channels unless stated otherwise. Furthermore, we assume equiprobable codewords: $\forall i \in \mathcal{M} : P_{X^n=\mathbf{c}_i} = P_{X^n=\mathbf{c}_1}$.

\section{Previous work}\label{sec:prev_work}
In \cite{early_harq_schemes2}, Berardinelli \emph{et al.} use a bit error estimate to perform a \ac{HARQ} feedback prediction. However, the formula based on the definition of \acp{LLR} provided in \cite{early_harq_schemes2} is faulty and has been corrected in \cite{arxiv_eharq_journal}:
\begin{align}
    P_{\mathrm{E}_i} = \frac{1}{1 + \exp{|\Lambda_i}|}, i=1,2,...,mp\,,
\end{align}
where $m$ is the number of bits per channel use and
\begin{align}\label{eq:llr}
{\Lambda}_i(y_{n(i)}) := \log\frac{P_{Y|b_i = 1}(y_{n(i)})}{P_{Y|b_i = 0}(y_{n(i)})},\, i=1,...,mp,
\end{align}
where $n(i)$ is the channel index associated with the $i$-th bit. The authors of \cite{early_harq_schemes2} further calculate the average of the bit error estimates and empirically determine a linear function to a pregiven threshold for the prediction. Instead of a linear function, we can also empirically determine a threshold, which yields the same decision regions. Then, the decision function results to
\begin{align}\label{eq:test_e}
    Z_\mathrm{E}(\mathbf{y}_p) := \begin{cases}
        1,&\text{if }(\frac{1}{mP}\sum_{i=1}^{mP}P_{\mathrm{E}_i})^{-1} \geq c_\mathrm{E}(\alpha)\\
        0,&\text{otherwise}
    \end{cases},
\end{align}
where $c_E(\alpha)$ is an empirically determined constant for a certain $\alpha$-level.

In \cite{ldpc_subcodes}, the authors extend this approach by applying a belief propagation decoder to the \acp{LLR} before calculating the bit error estimate. The bit error estimate is then 
\begin{align}
    P_{\mathrm{S}_i} = \frac{1}{1 + \exp{|\Lambda_i^{(l)}|}}, i=1,2,...,mP\,,
\end{align}
where $\Lambda_i^{(l)}$ is the a posteriori \ac{LLR} after $l$ iterations. The corresponding test $Z_{\mathrm{S}}$ is given analogously to (\ref{eq:test_e}) replacing $P_{\mathrm{E}_i}$ by $P_{\mathrm{S}_i}$.

In \cite{5961844}, Polyanskiy \emph{et al.} use in their achievability proof a mutual information density criterion to determine that sufficient redundancy has been received. The corresponding test for this predictor is given by
\begin{align}
    Z_{\mathrm{MI},\mathcal{I}}(\mathbf{y}_p) := \begin{cases}
        1,&\text{if }\max_{i \in \mathcal{I}}i_p(\mathbf{c}_i,\mathbf{y}_p) \geq c_\mathrm{MI}(\alpha)\\
        0,&\text{otherwise}
    \end{cases}.
\end{align}
With some abuse of notation, the mutual information density is defined as
\begin{align}
    i_p(\mathbf{c}, \mathbf{y}) := \log\frac{dP_{Y^p|X^n=\mathbf{c}}}{dP_{Y^p_{\mathrm{x}}}}(\mathbf{y}),
\end{align}
where $P_{Y^p_{\mathrm{x}}}$ is the output distribution generated by the full codebook $\mathcal{C}$, and $\mathcal{I} \in \{ \mathcal{M}_\mathrm{x}, \mathcal{M}\}$. The test in \cite{5961844} uses the first choice $\mathcal{I} = \mathcal{M}_\mathrm{x}$ due to the fact that the authors did not consider the presence of error detection codes explicitly. Hence, we designate the latter choice of $\mathcal{I} = \mathcal{M}$ as the error-detection-adapted version of Polyanskiy's decodability test, both of which are investigated in this work.

\section{Analytical results}
As Th.~\ref{th:np} suggests, the performance of \ac{HARQ} prediction schemes highly depend on the actual code and the used decoder. Additional to the previously studied \ac{HARQ} prediction schemes, classical decoding of the partially received signal vector, i.e. using $p$ symbols to perform decoding on them, also can be interpreted as a decodability prediction at a fixed $\alpha$-level. This scheme is commonly referred to as stop-feedback codes or proactive \ac{HARQ}. Here, an interesting question is, whether these schemes attain the minimum in (\ref{eq:np}). Before providing our first results, we first start with some definitions.
\begin{definition}{($D_p$-based prediction)}\label{def:dp}
A decoder $D_p : \mathcal{B}^p \mapsto \mathcal{M}_\mathrm{x}$ with a corresponding deterministic decoding region $\mathcal{R}_\mathrm{d} \subseteq \mathcal{B}^p$ and a randomized decoding region $\mathcal{R}_\mathrm{r} := \mathcal{B}^p \setminus \mathcal{R}_\mathrm{d}$ and the decision function
    \begin{align}
        Z_{D_p}(\mathbf{y}_p) := \begin{cases}1&\text{if }D_p(\mathbf{y}_p) \in \mathcal{M}\\0&\text{if }D_p(\mathbf{y}_p) \notin \mathcal{M} \end{cases}
    \end{align}
    is designated as $D_p$-based prediction.
\end{definition}
\begin{definition}{(Quasi-deterministic decoding)}
A decoder $D_p : \mathcal{B}^p \mapsto \mathcal{M}_\mathrm{x}$ with deterministic decoding region $\mathcal{R}_\mathrm{d}$ and randomized decoding region $\mathcal{R}_\mathrm{r}$, such that $\mathcal{R}_\mathrm{d} \cup \mathcal{R}_\mathrm{r} = \mathcal{B}^p$, $\mathcal{R}_\mathrm{d} \cap \mathcal{R}_\mathrm{r} = \emptyset$, and $P_{Y^p}(\mathcal{R}_\mathrm{r}) = 0$, is designated as a quasi-deterministic decoder.
\end{definition}
It is easy to show the existence of a decoder $D_p$ such that the associated decoding-based prediction is most powerful. First, we consider the Neyman-Pearson test from Th.~\ref{th:np} with $\tau = 1$. Then, we fix an $\alpha$-level together with the associated threshold $c$. Any decoder $D_p$ that outputs any value $D_p \in \mathcal{M}$ whenever the received signal vector $\mathbf{y}_p$ is in the acceptance region of this Neyman-Pearson test and $D_p \notin \mathcal{M}$ otherwise, is most powerful for the fixed $\alpha$-level. However, the question whether commonly used decoders are most powerful is usually hard to answer, as it also depends on the code itself. Nevertheless, we can consider the case $n = p$. The following proposition states an expected result that the $D$-based prediction with $D$ being a quasi-deterministic decoder is most powerful to predict the decoding outcome of itself.
\begin{prop}
    If $n=p$, the test $Z_{D_n}$, where $D_n = D$ is a quasi-deterministic decoder applied to $Y^n$, is most powerful for a certain $\alpha$-level.
\end{prop}
\ifarxiv
\begin{proof}
    See App.~\ref{app:prop1}.
\end{proof}
\fi

\subsection{Maximum-Likelihood Decoding}
\ac{ML} decoding is a widely considered decoding scheme, whose decoding complexity scales exponentially with the information length in general. From theoretic perspective, \ac{ML} decoding can be well analyzed and is given by
\begin{align}\label{eq:infodens}
    D_\mathrm{ML}(\mathbf{y}_n) := \argmax_i i_n(\mathbf{c}_i; \mathbf{y}_n)\,.
\end{align}
The following theorem gives the conditional correct decoding probability. By applying it to Th.~\ref{th:np}, we can define the most powerful test.
\begin{thrm}
For an arbitrary code $\mathcal{C}$ with corresponding \ac{ML} decoding, the conditional correct decoding probability $P_{D|Y^pX^n}$ is given by
    \begin{align}\label{eq:pd_ml}
        P_{D|Y^pX^n}&(i|\mathbf{y}_p, \mathbf{c}_i) = \sum_{l=0}^{|\mathcal{M}_i|}\frac{1}{l+1} \sum_{\mathcal{G} \in {\mathcal{M}_i \choose l}}\nonumber\\
        P_{Y^r|X^n=\mathbf{c}_i}&\big[\{{\cap_{j\in \mathcal{M}_i\setminus \mathcal{G}}\{l(\mathbf{c}_i, \mathbf{c}_j;\mathbf{y}_p) > l(\mathbf{c}_j, \mathbf{c}_i;Y^r)\}}\}\nonumber\\
        \bigcap &\{\cap_{j\in \mathcal{G}}\{l(\mathbf{c}_i, \mathbf{c}_j;\mathbf{y}_p) = l(\mathbf{c}_j, \mathbf{c}_i;Y^r)\}\}\big]\,,
    \end{align}
    where $\mathcal{M}_i := \mathcal{M}_\mathrm{x}\setminus \{i\}$ and $l(\mathbf{x}_1, \mathbf{x}_2; \mathbf{y}) := \log \frac{dP_{Y|X=\mathbf{x}_1}}{dP_{Y|X=\mathbf{x}_2}}(\mathbf{y})$.
\end{thrm}
\ifarxiv
\begin{proof}
    See App.~\ref{app:th2}.
\end{proof}
\fi
For the \ac{AWGN} channel with the noise distribution given by $\mathcal{N}(\mathbf{0}, \sigma^2 \mathbf{I}_n)$, we can particularize the decoding probability in (\ref{eq:pd_ml}) to the cumulative distribution function of a multivariate normal distribution.
\begin{cor}\label{th:decawgn}
    For an \ac{AWGN} channel, the decodability distribution of \ac{ML} decoding reduces to
    \begin{align}\label{eq:pd_awgn}
        P_{D|Y^pX^n}(i|\mathbf{y}_p,\mathbf{c}_i) = \Phi_{|\mathcal{M}_\mathrm{x}|}\Big[ \mathbf{d} - \mathbf{B}_{p|i}(\mathbf{y}_p - \mathbf{c}_i) \Big]\,,
    \end{align}
    where $\Phi$ is the cumulative distribution function of the $|\mathcal{M}_\mathrm{x}|$-variate normal distribution with $\mu = \mathbf{0}$ and $\Sigma = \sigma^2 \mathbf{B}_{r|i}\mathbf{B}_{r|i}^\top$, and
    \begin{align}
        \mathbf{B}_i := \begin{pmatrix}
                    \mathbf{c}_1^\top - \mathbf{c}_i^\top\\
                    \vdots\\
                    \mathbf{c}_{|\mathcal{M}_\mathrm{x}|}^\top-\mathbf{c}_i^\top
                \end{pmatrix} \in \mathbb{R}^{|\mathcal{M}_\mathrm{x}| \times n}
    \end{align}
    with $\mathbf{B}_{p|i} := \mathbf{B}_i[1,...,|\mathcal{M}_\mathrm{x}|;1,...,p]$, $\mathbf{B}_{r|i} := \mathbf{B}_i[1,...,|\mathcal{M}_\mathrm{x}|;p+1,...,n]$, and $\mathbf{d} := (||\mathbf{c}_1||^2,...,||\mathbf{c}_{|\mathcal{M}_\mathrm{x}|}||^2)^\top$.
\end{cor}
\ifarxiv
\begin{proof}
    See App.~\ref{app:cor1}.
\end{proof}
\fi
Cor.~\ref{th:decawgn} shows that in an \ac{AWGN} channel with \ac{ML} decoding, the successful partial signal vectors $(Y^p|{\mathcal{D}_\mathrm{A}X^n=\mathbf{c}_i}) \sim GSN_{p,|\mathcal{M}_\mathrm{x}|}(\mathbf{c}_i, \sigma^2\mathbf{I}_p,-\mathbf{B}_{p|i},-\mathbf{d}-\mathbf{B}_{p|i}\mathbf{c}_i,\sigma^2\mathbf{B}_{r|i}\mathbf{B}_{r|i}^\top)$ are distributed according to a general skew normal distribution with $|\mathcal{M}_\mathrm{x}|$ dimensions, as defined by Gupta et. al. in \cite{GUPTA2004181}. Unfortunately, the evaluation of the probability density function of this distribution is computationally complex. Nevertheless, for numerical evaluation, we can reduce the dimensionality by removing redundancies in the multivariate skew normal distribution. In particular, rows of $\mathbf{B}_{r|i}$ that are equal can be collapsed into a single row, which reduces the dimensionality of the distribution.\\

Another interesting case are classical stop-feedback codes. In this scenario, the block length and prediction length is equal, $n=p$. We have already shown that the decoding-based prediction using the same decoder as the actual decoder is most powerful for a certain $\alpha$-level. For the mutual information density prediction under the absence of error detection, we can provide an even stronger statement for quasi-deterministic \ac{ML} decoding.
\begin{prop}\label{prop:mi-crc}
    For the case of $n=p$ and $\mathcal{M} = \mathcal{M}_\mathrm{x}$ with quasi-deterministic \ac{ML} decoding, the test $Z_{\mathrm{MI},\mathcal{M}}$ is most powerful for any achievable $\alpha$-level, if the correct decoding probabilities $P_{D|X^n}(i|\mathbf{c}_i)$ of all codewords are equal.
\end{prop}
\ifarxiv
\begin{proof}
    See App.~\ref{app:prop2}.
\end{proof}
\fi
This result is particularly interesting as it shows that for the transmission scenario without error detection and codebooks that have symmetric \ac{ML} decoding regions, the test described in \cite{5961844} is most powerful. Note that due to the non-randomized nature of this test, not all $\alpha$-levels may be achievable. Nevertheless, by introducing randomization into $Z_{\mathrm{MI},\mathcal{M}}$ for the case of equality, we can extend Prop.~\ref{prop:mi-crc} to all $\alpha$-levels.

\subsection{t-error correcting codes under hard-decision decoding}
Algebraic block codes, such as Reed-Solomon codes, are usually decoded using hard-decision decoding. Hard-decision decoding involves quantizing the received signal to discrete values. The following proposition gives us the conditional correct decoding probability for $t$-error correcting codes.
\begin{thrm}
    For an arbitrary $t$-error-correcting code under a corresponding hard-decision decoder, and a symbol-wise quantization function $\mathbf{Q}$ with equal elements $Q : \mathcal{B} \to \mathbb{F}_2^m$, the decoding probability yields
    \begin{align}\label{eq:terrpd}
        P_{D|Y^pX^n}(i|\mathbf{y}_p,\mathbf{c}_i) = \sum_{u=0}^{t-d_\mathrm{H}(\mathbf{c}_i,\mathbf{Q}(\mathbf{y}_p))} {n-p \choose u}v^{u}(1-v)^{n-p-u}\,,
    \end{align}
    where $v$ is the associated quantization bit error probability and $d_\mathrm{H}$ is the Hamming distance.
\end{thrm}
\ifarxiv
\begin{proof}
    See App.~\ref{app:th3}.
\end{proof}
\fi
We note that the conditional correct decoding probability only depends on the previously received signal vector in terms of the already "occured" bit errors. This allows us to derive a different representation for linear codes. Linear codes can further be represented by their parity-check matrix. In this case, we can further establish a link to the syndrome coset of the code.
\begin{thrm}\label{th:t_parity}
    Suppose that $\forall i \in \mathcal{M} : P_{D|X^n}(i|\mathbf{c}_i) = P_{\mathcal{D}_\mathrm{A}}$. For a linear $t$-error-correcting code under a corresponding hard-decision decoder, where the code is described by its parity-check matrix $\mathbf{H}$, the Neyman-Pearson quotient is given by
    \begin{align}
    T_\mathrm{NP}(\mathbf{y}_p) = \frac{1}{P_{\mathcal{D}_\mathrm{A}}}{\sum_{\mathbf{e} \in \mathcal{C}(\mathbf{s})} P_{E^p|Y^p}(\mathbf{e}|\mathbf{y}_p) P_{\mathcal{D}_\mathrm{A}|E^p=\mathbf{e}}}\,,
    \end{align}
    where $\mathbf{s} := \mathbf{H}_p\mathbf{Q}(\mathbf{y}_p)$ is the syndrome and $\mathcal{C}(\mathbf{s})$ is the coset associated with syndrome $\mathbf{s}$.
\end{thrm}
\ifarxiv
\begin{proof}
    See App.~\ref{app:th4}.
\end{proof}
\fi
This theorem allows us to establish a theoretic categorization of the \ac{LLR}-based test $Z_\mathrm{E}$ proposed in \cite{early_harq_schemes2}.
\begin{prop}
    Let $\mathcal{A}^n = \mathbb{F}_2^n$ and suppose that the distribution of \acp{LLR} is i.i.d. Then, $Z_\mathrm{E}$ is most powerful for any achievable $\alpha$-level for a code that contains all $\mathcal{C} = \mathbb{F}_2^n$, if the \acp{LLR} follow a Dirac delta distribution.
\end{prop}
\ifarxiv
\begin{proof}
    See App.~\ref{app:prop3}.
\end{proof}
\fi
Th.~\ref{th:t_parity} allows us to interpret the \ac{LLR} test $Z_E$ as the most powerful test for a $0$-error correcting code containing all binary sequences in its codebook. Although the theorem requires that the \acp{LLR} are distributed according to a Dirac delta distribution, this property is also approximately fulfilled for very small bit error probabilities i.e. high \ac{SNR} in an \ac{BI-AWGN} channel.



\subsection{Sum-product decoder}\label{sec:sp}
The sum-product decoder is an iterative decoder, which is known to be the bitwise \ac{MAP} decoder on cycle-free tree graphs. However, modern \ac{LDPC} codes contain cycles, which makes an exact analytical analysis of the performance on these codes a hard problem. In particular, the decoding probability $P_{D|Y^pX^n}$ cannot be stated in a closed form. Hence, we approximate $dP_{Y^p|\mathcal{D}_\mathrm{A}}$ numerically by using a kernel density estimation with normal kernels. However, with increasing $p$ also the dimensionality of the approximated probability density increases, which requires significantly larger sample sizes \cite{kde_book}. On top of that, $dP_{Y^p|\mathcal{D}_\mathrm{A}}$ is a mixed distribution with $|\mathcal{M}|$ components. This increases the difficulty of applying a kernel density estimation even further. Hence, we define a measurable function $\mathbf{f}_i : \mathbf{y}_p \mapsto (\mathbf{y}_p - \mathbf{c}_i + \mathbf{c}_1)$. Under the assumption of symmetry with respect to $P_{Y^p|X^p}$, which clearly is given for an \ac{BI-AWGN} channel, the probability density becomes
\begin{align}
dP_{Y^p|\mathcal{D}_\mathrm{A}} = \sum_{i \in \mathcal{M}}\frac{ dP_{Y^p|X^n}(\mathbf{f}_i(\mathbf{y}_p) | \mathbf{c}_1) P_{D|Y^pX^n}(1|\mathbf{f}_i(\mathbf{y}_p), \mathbf{c}_1)}{|\mathcal{M}| P_{D|X^n}(i|\mathbf{c}_i)}\,.
\end{align}
Hence, the estimation task is reduced to estimating only a single component instead of the whole mixed distribution.


\section{Numerical results}
In this section, we present our results for a very short blocklength of $n=24$ in a \ac{BI-AWGN} channel at an \ac{SNR} of 5~dB. Due to the increasing evaluation complexity of the derived terms, we restrict to such a short blocklength. For all the decoding schemes, we sample data using Monte-Carlo simulations and apply the respective tests to the samples. For the sum-product decoder, we approximate the uniformly most powerful test using a kernel-density estimation, as described in Sec.~\ref{sec:sp}. We use a (3,6)-regular \ac{LDPC} code with information length $k=14$. For error detection, we employ a 4~bit CRC with the polynomial $1 + x^2 + x^3 + x^4$.\\

\begin{figure}
    \centering
    \includegraphics[width=0.9\columnwidth]{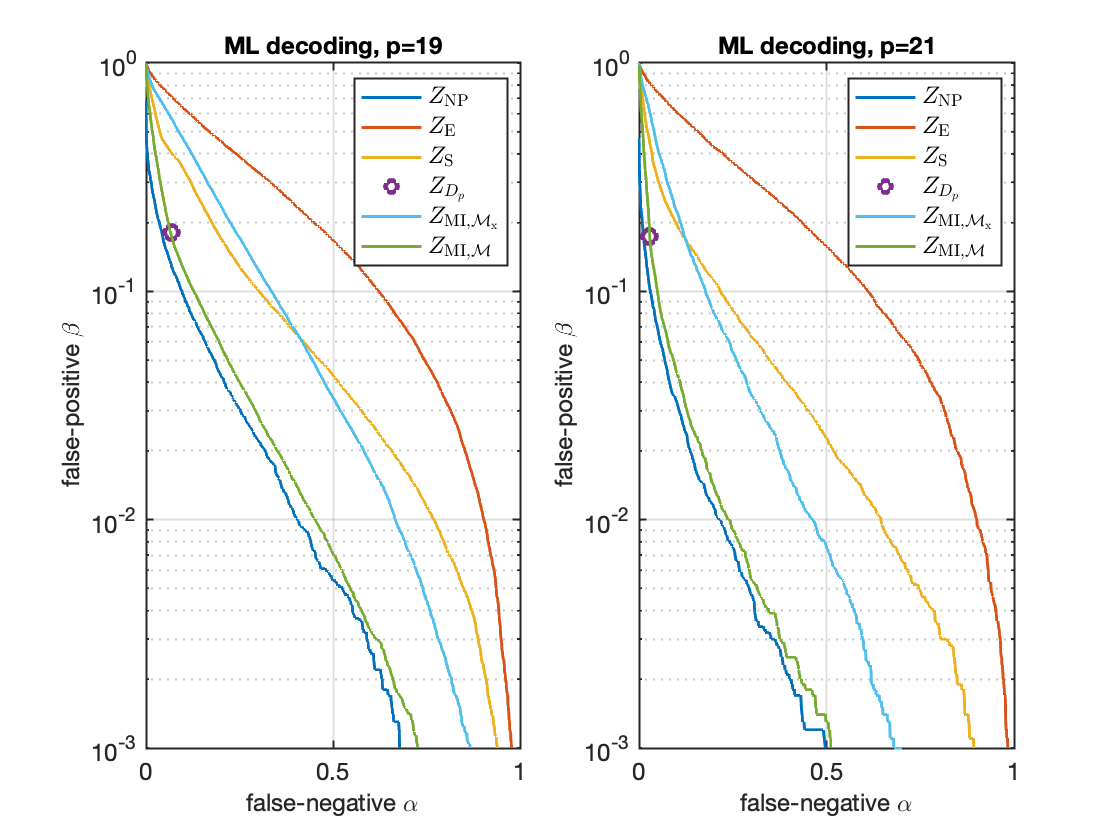}
    \caption{HARQ prediction performance for a (3,6)-regular \ac{LDPC} code with $n=24$ and $k=14$ under maximum-likelihood decoding in the BI-AWGN channel at an \ac{SNR} of 5~dB with prediction length $p=19$ and $p=21$.}
    \label{fig:awgn_snr5_ml}
\end{figure}
Fig.~\ref{fig:awgn_snr5_ml} shows the false-positive $\beta$ performance over the false-negative $\alpha$ rate of a (3,6)-regular \ac{LDPC} code under \ac{ML} decoding with information length $k=14$ and prediction length $p=19$, $p=21$. We observe that the error-detection-adapted mutual information test $Z_{\mathrm{MI},\mathcal{M}}$ achieves the closest performance to the Neyman-Pearson test. However, there is still a clear gap between both. Furthermore, we note that the decoding-based prediction achieves only a certain work point of $Z_{\mathrm{MI},\mathcal{M}}$. Furthermore, we see that the subcode-based prediction $Z_\mathrm{S}$ and the error-detection-adapted mutual information test $Z_{\mathrm{MI},\mathcal{M}_\mathrm{x}}$ reach a comparable performance, where $Z_{\mathrm{MI},\mathcal{M}_\mathrm{x}}$ performs slightly worse at small $\alpha$, however outperforms $Z_\mathrm{S}$ at larger $\alpha$. Clearly, the \ac{LLR}-based prediction $Z_\mathrm{E}$ performs the worst. 
\begin{figure}
    \centering
    \includegraphics[width=0.9\columnwidth]{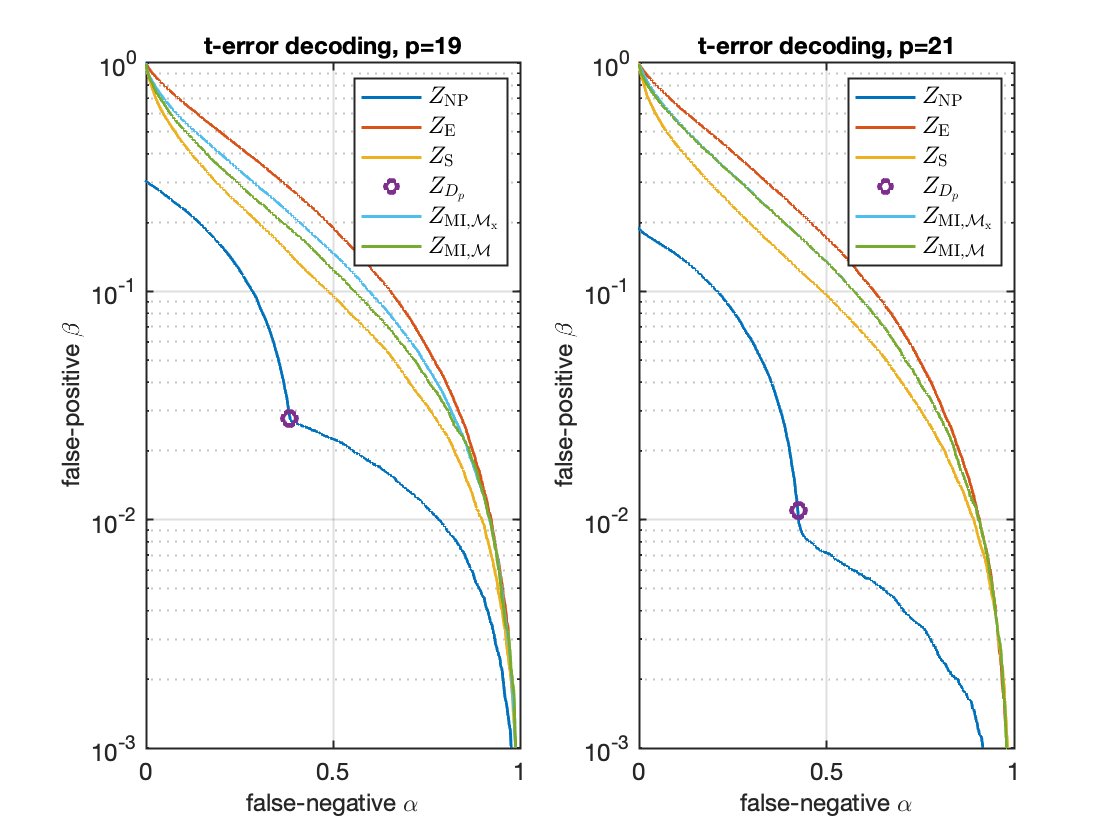}
    \caption{HARQ prediction performance for a 1-error correcting code with $n=24$ and $k=14$ under hard decision decoding in the BI-AWGN channel at an \ac{SNR} of 5~dB with prediction length $p=19$ and $p=21$.}
    \label{fig:awgn_snr5_t}
\end{figure}
For $t$-error decoding, in Fig.~\ref{fig:awgn_snr5_t}, we see that the decoding-based prediction achieves a working point of the Neyman-Pearson test. For the Neyman-Pearson test $Z_\mathrm{NP}$, we observe two linear regions corresponding to allowing up to 1 bit error and 0 bit errors in $\mathbf{y}_p$, respectively. Among the other prediction schemes, which show a significant gap to the Neyman-Pearson test, the subcode-based prediction $Z_\mathrm{S}$ performs the best.
\begin{figure}
    \centering
    \includegraphics[width=0.9\columnwidth]{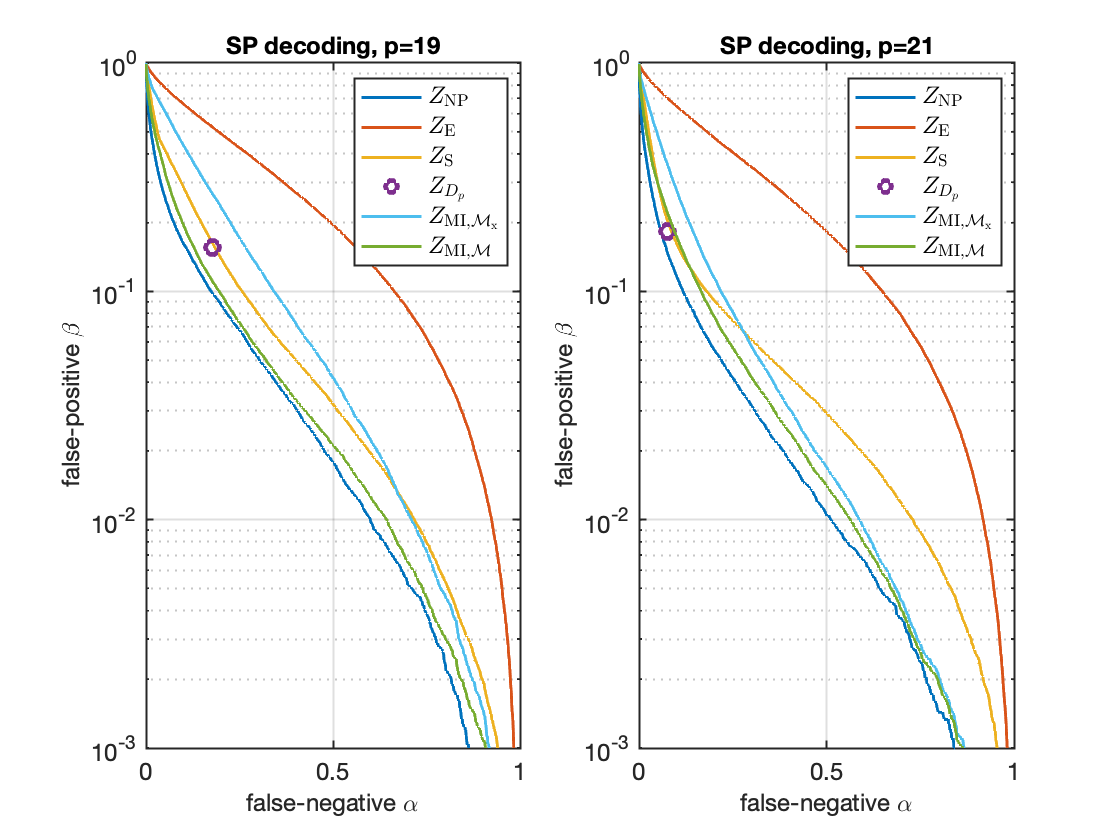}
    \caption{HARQ prediction performance for a (3,6)-regular \ac{LDPC} code with $n=24$ and $k=14$ under sum-product decoding in the BI-AWGN channel at an \ac{SNR} of 5~dB with prediction length $p=19$ and $p=21$.}
    \label{fig:awgn_snr5_sp}
\end{figure}
For the sum-product decoder, in Fig.~\ref{fig:awgn_snr5_sp}, we notice that the decoding-based prediction achieves a working point only slightly better than the subcode-based prediction indicating that the additional iterations do not contribute much in terms of prediction power. As for the \ac{ML} decoding, the error-detection-adapted mutual information test $Z_{\mathrm{MI},\mathcal{M}}$ comes close to the performance of the Neyman-Pearson test.

For all decoder types, we note that the Neyman-Pearson test achieves a better performance for a larger prediction length $p$. The same is true also for the other tests. However, the improvement from the longer prediction length is not the same for all tests. Especially, the subcode test $Z_\mathrm{S}$ benefits the most except for $t$-error decoding. For $t$-error decoding, only the decoding-based prediction $Z_{D_p}$ improves with a larger prediction length.

\section{Conclusions}
In this work, we have derived the Neyman-Pearson tests for maximum-likelihood decoding and $t$-error codes. We particularized and applied the Neyman-Pearson test to the \ac{BI-AWGN} channel. Furthermore, we showed in numerical evaluations that none of the previously studied prediction schemes is most powerful for maximum-likelihood decoding and sum-product decoding. In particular, we demonstrate that decoding and estimating the decodability, i.e. \ac{HARQ} prediction, are different problems in general.

\bibliographystyle{IEEEtran}
\bibliography{lib}

\ifarxiv
\newpage
\appendices
\section{Proof of Theorem 1}\label{app:th1}
\begin{proof}
An optimal test using the probability densities ratio is given by the Neyman-Pearson lemma:
\begin{align}
    Z(\mathbf{y}_p) = \begin{cases}
                1,&\text{if }T_\mathrm{NP}(\mathbf{y}_p) > \gamma\\
                Z_\tau,&\text{if }T_\mathrm{NP}(\mathbf{y}_p) = \gamma\\
                0,&\text{if }T_\mathrm{NP}(\mathbf{y}_p) < \gamma\\
            \end{cases}\,,
\end{align}
with $T_\mathrm{NP}(\mathbf{y}_p) := \frac{dP_{Y^p|\mathcal{D}_\mathrm{A}}(\mathbf{y}_p)}{dP_{Y^p|\mathcal{D}_\mathrm{N}}(\mathbf{y}_p)}$ and $\gamma > 0$, where the randomization at $\gamma$ may be required to reach certain $\alpha$-levels. Then, we can equivalently transform the acceptance region of the Neyman-Pearson test as
\begin{align}
    \Bigg\{\frac{dP_{Y^p|\mathcal{D}_\mathrm{A}}(\mathbf{y}_p)}{dP_{Y^p|\mathcal{D}_\mathrm{N}}(\mathbf{y}_p)} &> \gamma\Bigg\} = \Bigg\{\frac{dP_{Y^p|\mathcal{D}_\mathrm{N}}(\mathbf{y}_p)}{dP_{Y^p|\mathcal{D}_\mathrm{A}}(\mathbf{y}_p)} &< \frac{1}{\gamma}\Bigg\}\\
    = \Bigg\{\frac{dP_{Y^p}(\mathbf{y}_p)}{dP_{Y^p|\mathcal{D}_\mathrm{A}}(\mathbf{y}_p)} &< P_{\mathcal{D}_\mathrm{N}}\frac{1}{\gamma} + P_{\mathcal{D}_\mathrm{A}}\Bigg\}\\
    = \Bigg\{\frac{dP_{Y^p|\mathcal{D}_\mathrm{A}}(\mathbf{y}_p)}{dP_{Y^p}(\mathbf{y}_p)} &> \frac{1}{P_{\mathcal{D}_\mathrm{N}}\frac{1}{\gamma} + P_{\mathcal{D}_\mathrm{A}}}\Bigg\}\,.
\end{align}
Obviously, the same transforms can be applied analogously to the randomized and the rejection region of the test, respectively. The output distribution of the channel is generated by the valid codewords of the code:
\begin{align}
    dP_{Y^p} &= \frac{1}{|\mathcal{M}|}\sum_{i \in \mathcal{M}} dP_{Y^p|X^n}(\mathbf{y}_p | \mathbf{c}_i)\,.
\end{align}
The distribution of decodable codewords depends on the valid codewords and the decoder:
\begin{align}
    dP_{Y^p|\mathcal{D}_\mathrm{A}} &= \frac{1}{|\mathcal{M}|}\sum_{i \in \mathcal{M}} dP_{Y^p|X^n\mathcal{D}_\mathrm{A}}(\mathbf{y}_p | \mathbf{c}_i)\\
    &= \frac{1}{|\mathcal{M}|}\sum_{i \in \mathcal{M}} \frac{dP_{Y^p|X^n}(\mathbf{y}_p | \mathbf{c}_i)P_{\mathcal{D}_\mathrm{A}|Y^pX^n}(\mathbf{y}_p, \mathbf{c}_i)}{P_{\mathcal{D}_\mathrm{A}|X^n = \mathbf{c}_i}}\label{eq:pda2}\\
    &= \frac{1}{|\mathcal{M}|}\sum_{i \in \mathcal{M}} \frac{dP_{Y^p|X^n}(\mathbf{y}_p | \mathbf{c}_i)P_{D|Y^pX^n}(i|\mathbf{y}_p, \mathbf{c}_i)}{P_{D|X^n}(i|\mathbf{c}_i)}\label{eq:pda3}\,,
\end{align}
where (\ref{eq:pda2}) results from Bayes and (\ref{eq:pda3}) is the conditional correct decoding probability, which can be obtained by inserting the actual decoder.

\end{proof}

\section{Proof of Proposition 1}\label{app:prop1}
For a quasi-deterministic decoder, we only have to consider the deterministic region, because any property that is $\mu$-a.s. valid for the deterministic region $\mathcal{R}_\mathrm{d}$, is also $\mu$-a.s. valid for the whole decoding region $\mathcal{B}^n$.
$Z_{D_n}$ only takes values in $\{0,1\}$. Hence, there exists an $\alpha \in [0,1]$ such that 
\begin{align}
    \mathbb{E}[Z_{D_n}|\mathcal{D}_\mathrm{A}] = 1 - \alpha\,.
\end{align}
This is the $\alpha$-level of the test $Z_{D_n}$. Furthermore, for any decoder, we can state that there exists a $\delta > 0$, such that 
\begin{align}
    \essinf_{\mathbf{y}_n \in \mathcal{R}_\mathrm{d} : D_n(\mathbf{y}_n) \in \mathcal{M}} \frac{dP_{Y^n|\mathcal{D}_\mathrm{A}}}{dP_{Y^n}} = \delta.
\end{align}
Otherwise, there would exist a set $\mathcal{Y} \subset \mathcal{R}_\mathrm{d}$ with $P_{Y^n}(\mathcal{Y}) > 0$, such that for all $\mathbf{y}_n \in \mathcal{Y}$ there exists an $i \in \mathcal{M}$ with $D_n = i$ and $P_{D|Y^nX^n}(i|\mathbf{y}_n,\mathbf{c}_i) = 0$. Obviously, this is a contradiction due to $D_n = D$. Analogously, we can also show that the following equation holds
\begin{align}
    \esssup_{\mathbf{y}_n \in \mathcal{R}_\mathrm{d} : D_n(\mathbf{y}_n) \notin \mathcal{M}} \frac{dP_{Y^n|\mathcal{D}_\mathrm{A}}}{dP_{Y^n}} = 0.
\end{align}
Hence, the test $Z_{D_n}$ behaves $\mu$-a.s. as the $\alpha$-level Neyman-Pearson test with $\tau = 1$ and $\gamma = \delta$ and therefore, is most powerful by the fundamental lemma of Neyman and Pearson\cite[Th.~3.2.1]{lehmann2005testing}.

\section{Proof of Theorem 2}\label{app:th2}
\begin{proof}
As stated in (\ref{eq:infodens}), the maximum-likelihood decoder chooses the codeword that maximizes the mutual information density. In the case that multiple codewords have the same maximal information density, the decoder chooses one of these codewords randomly. Because we assume a memoryless channel, the information density can be split into an already determined and a random part with $\mathbf{y}_p \in \mathcal{B}^p$ and $\mathbf{y}_r \in \mathcal{B}^{n-p}$:
\begin{align}
    i_n(\mathbf{c}_i; [\mathbf{y}_p,\mathbf{y}_r]) = \log\frac{dP_{Y^p|X^p}}{dP_{Y^p_\mathrm{x}}}(\mathbf{y}_p) + \log\frac{dP_{Y^r|X^r}}{dP_{Y^r_\mathrm{x}}}(\mathbf{y}_r)\,,
\end{align} 
where $P_{Y^p_\mathrm{x}}$ and $P_{Y^r_\mathrm{x}}$ designate the respective output distribution generated by the full codebook $\mathcal{C}$ with equiprobable codewords.
Hence, the correct decoding probability is the probability that given $i(\mathbf{c}_i; \mathbf{y}_p)$, $i(\mathbf{c}_i; (\mathbf{y}_p,\mathbf{y}_r))$ either exclusively achieves the highest information density among all codewords or together with $l$ further codewords. In the latter case the decoder chooses the codeword randomly. With these thoughts, the decoding probability results to
\begin{align}
    P_{D|Y^pX^n}&(i|\mathbf{y}_p, \mathbf{c}_i) = \sum_{l=0}^{|\mathcal{M}_i|}\frac{1}{l+1} \sum_{\mathcal{G} \in {\mathcal{M}_i \choose l}}\nonumber\\
        P_{Y^r|X^n=\mathbf{c}_i}&\big[\{\cap_{j\in \mathcal{M}_{i}\setminus \mathcal{G}}\{i_n(\mathbf{c}_i;[\mathbf{y}_p, Y^r]) > i_n(\mathbf{c}_j;[\mathbf{y}_p,Y^r])\}\}\nonumber\\
        \bigcap &\{\cap_{j\in \mathcal{G}}\{i_n(\mathbf{c}_i;[\mathbf{y}_p,Y^r]) = i_n(\mathbf{c}_j;[\mathbf{y}_p,Y^r])\}\}\big]\,,
\end{align}
where $\mathcal{M}_i := \mathcal{M}_{\mathrm{x}} \setminus \{i\}$. Together with the definition of $l(\cdot)$ we get the proposition.
\end{proof}

\section{Proof of Corollary 1}\label{app:cor1}
\begin{proof}
    The probability density with respect to the Lebesgue measure of the \ac{AWGN} channel is a continuous function over the domain of $\mathbb{R}^n$. Hence, the probability that the mutual information densities are equal simply results to zero. This reduces the decoding probability to 
    \begin{align}\label{eq:pd_awgn1}
         P&_{D|Y^pX^n}(i|\mathbf{y}_p, \mathbf{c}_i)\nonumber\\
        =&P_{Y^r|X^n=\mathbf{c}_i}\big[\cap_{j\in \mathcal{M}_\mathrm{x}\setminus \{i\}}\{l(\mathbf{c}_i,\mathbf{c}_j;\mathbf{y}_p) > l(\mathbf{c}_j,\mathbf{c}_i;Y^r)\}\big]\,.
    \end{align}
    Under $P_{Y^r|X^n=\mathbf{c}_i}$, $l(\mathbf{c}_j,\mathbf{c}_i;Y^r)$ is distributed as
    \begin{align}\label{eq:lr}
        L_r := \frac{1}{2\sigma^2}\log{e}\sum_{k=p+1}^n 2(c_{i,k} - c_{j,k})Z_k + (c_{i,k} - c_{j,k})^2\,,
    \end{align}
    where $c_{i,k}$ is the $k$-th component of $\mathbf{c}_i$, $Z_k \sim \mathcal{N}(0,\sigma^2)$ are i.i.d. Furthermore, the predetermined part of the mutual information density is given by
    \begin{align}\label{eq:lp}
        l(\mathbf{c}_i,\mathbf{c}_j;\mathbf{y}_p)& = \frac{1}{2\sigma^2}\log{e}\nonumber\\
        &\cdot\sum_{k=p+1}^n 2(c_{j,k} - c_{i,k})(y_{p,k}-c_{i,k}) + (c_{i,k} - c_{j,k})^2\,,
    \end{align}
    where $y_{p,k}$ is the $k$-th component of $\mathbf{y}_p$.
    Inserting (\ref{eq:lr}) and (\ref{eq:lp}) into (\ref{eq:pd_awgn1}) gives the desired statement.
\end{proof}

\section{Proof of Proposition 2}\label{app:prop2}
\begin{proof}
    For a quasi-deterministic decoder $D$, any property that is $\mu$-a.s. valid for the deterministic region, is also $\mu$-a.s. valid for the whole region $\mathcal{B}^n$. On the deterministic region $\mathcal{R}_\mathrm{d}$, the functions $P_{D=i|Y^pX^n=\mathbf{c}_i} : \mathcal{B}^p \to [0,1], i \in \mathcal{M},$ are indicator functions with mutually exclusive regions on $\mathcal{B}^p$. We first assume that the correct decoding probabilities are equal: $\forall i \in \mathcal{M} : P_{D|X^n}(i|\mathbf{c}_i) = P_{D|X^n}(i_0|\mathbf{c}_{i_0})$, where $i_0 \in \mathcal{M}$. Hence, for all $\mathbf{y}_n : D(\mathbf{y}_n) \in \mathcal{M}$, the Neyman-Pearson quotient reduces to
    \begin{align}
        T_\mathrm{NP}(\mathbf{y}_n) &= \frac{dP_{Y^n|X^n}(\mathbf{y}_n|\mathbf{c}_{D(\mathbf{y_n})})}{dP_{Y^n}(\mathbf{y}_n)}\frac{1}{\mathcal{M}P_{D|X^n}(D(\mathbf{y}_n)|\mathbf{c}_{{D(\mathbf{y}_n)}})}\\
        &= \frac{\exp{\left(i_n(\mathbf{c}_{{D(\mathbf{y}_n)}},\mathbf{y}_n)\right)}}{\mathcal{M}P_{D|X^n}(i_0|\mathbf{c}_{i_0})}\,,\label{eq:np_2}
    \end{align}
    and $T_\mathrm{NP}(\mathbf{y}_n) = 0$ otherwise. Note that $P_{Y^n} = P_{Y^n_\mathrm{x}}$ due to $\mathcal{M} = \mathcal{M}_\mathrm{x}$. The mutual information density is always larger or equal to $0$. So, there exists a constant $\gamma_\mathrm{min} > 0$, such that $\essinf_{\mathbf{y}_n : D(\mathbf{y}_n) \in \mathcal{M}}T_\mathrm{NP}(\mathbf{y}_n) = \gamma_\mathrm{min}$. Because the \ac{ML} decoder always chooses the codeword that maximizes the mutual information density, we can derive an equivalent form of the Neyman-Pearson test with $\tau = 1$ for any $\gamma > \gamma_\mathrm{min}$:
    \begin{align}
        Z_\mathrm{NP} = \begin{cases}
            1,&\text{if } \max_{i\in\mathcal{M}}i_n(\mathbf{c}_i,\mathbf{y}_n) \geq \delta_\gamma\,,\\
            0,&\text{otherwise}\,,
        \end{cases}
    \end{align}
    with $\delta_\gamma := \log(\mathcal{M}P_{D|X^n}(i_0,\mathbf{c}_{i_0})\gamma)$, which is also the mutual information density test $Z_{\mathrm{MI},\mathcal{M}}$ with $c_\mathrm{MI} = \delta_\gamma$.
\end{proof}

\section{Proof of Theorem 3}\label{app:th3}
\begin{proof}
    Given that the signal $\mathbf{y}_p$ was received and $\mathbf{c}_i$ was transmitted, the number of already occured bit errors is the Hamming distance between the quantized signal vector and the codeword $k := d_\mathrm{H}(\mathbf{c}_i, \mathbf{Q}(\mathbf{y}_p)$. Hence, the correct decoding probability is the probability that, given $k$ bit errors occured already, the total number of bit errors does not exceed $t$.
\end{proof}

\section{Proof of Theorem 4}\label{app:th4}
\begin{proof}
    Let $\mathbf{H}_p$ be the parity-check matrix of the subcode, such that $\forall i \in \mathcal{M} : \mathbf{H}_p \mathbf{c}_i^p = 0 $. Note that $\mathbf{H}_p$ does not only include the parity-check constraints of the error correction code itself but also the check constraints of the error detection code, i.e. \ac{CRC}. The syndrome of the received signal vector $\mathbf{y}_p$ is given by $\mathbf{s} := \mathbf{H}_p \mathbf{Q}(\mathbf{y}_p)$. Let the coset of $\mathbf{s}$ be the set
    \begin{align}
        \mathcal{C}(\mathbf{s}) := \{ \mathbf{b}_p \in \mathbb{F}_2^p : \mathbf{H}_p \mathbf{b}_p = \mathbf{s}\}\,.
    \end{align}
    Furthermore, the following equality holds for all valid codewords
    \begin{align}
        \forall i \in \mathcal{M} : \mathbf{s} &= \mathbf{H}_p \mathbf{Q}(\mathbf{y}_p)= \mathbf{H}_p \mathbf{c}_i + \mathbf{H}_p (\mathbf{Q}(\mathbf{y}_p) \oplus \mathbf{c}_i)\\
        & = \mathbf{H}_p (\mathbf{e} := \mathbf{Q}(\mathbf{y}_p) \oplus \mathbf{c}_i)\,,
    \end{align}
    where $\mathbf{Q}(\mathbf{y}_p) \oplus \mathbf{c}_i$ is the element-wise addition on $\mathbb{F}_2$. Hence, we can conider the error pattern $\mathbf{e}$ instead of the particular codewords.  
    Furthermore, the decoding probability in (\ref{eq:terrpd}) depends only on $\mathbf{y}_p$ in terms of the Hamming distance. Hence, by replacing the Hamming distance $d_\mathrm{H}(\mathbf{c}_i, \mathbf{Q}(\mathbf{y}_p)) = d_\mathrm{H}(\mathbf{e}, \mathbf{0})$ by the Hamming distance of the error pattern to the all zeros vector, we obtain
    \begin{align}
        P_{D|Y^pX^n}&(i|\mathbf{y}_p,\mathbf{c}_i) = P_{\mathcal{D}_\mathrm{A}|E^p=\mathbf{e}}\\
        &= \sum_{u=0}^{t-d_\mathrm{H}(\mathbf{e},\mathbf{0})} {n-p \choose u}v^{u}(1-v)^{n-p-u}\,.
    \end{align}
    In addition, the probability of receiving a certain signal vector $\mathbf{y}_p$ is determined by the probability of the error pattern:
    \begin{align}
        &\sum_{i \in \mathcal{M}} P_{Y^p|X^p}(\mathbf{y}_p|\mathbf{c}_i) \frac{P_{D|Y^pX^n}(i|\mathbf{y}_p,\mathbf{c}_i)}{|\mathcal{M}|P_{D|X^n}(i|\mathbf{c}_i)}\\
        &= \sum_{i \in \mathcal{M}}\frac{P_{D|Y^pX^n}(i|\mathbf{y}_p,\mathbf{c}_i)}{|\mathcal{M}|P_{D|X^n}(i|\mathbf{c}_i)}\sum_{\mathbf{e} \in \mathcal{C}(\mathbf{s})} P_{Y^p|X^p}(\mathbf{y}_p|\mathbf{c}_i) P_{E^p|Y^p}(\mathbf{e}|\mathbf{y}_p)\label{eq:channelprob_2} \\
        &= \frac{1}{|\mathcal{M}|P_{\mathcal{D}_\mathrm{A}}}\sum_{\mathbf{e} \in \mathcal{C}(\mathbf{s})} P_{E^p|Y^p}(\mathbf{e}|\mathbf{y}_p)P_{\mathcal{D}_\mathrm{A}|E^p=\mathbf{e}} \sum_{i \in \mathcal{M}}P_{Y^p|X^p}(\mathbf{y}_p|\mathbf{c}_i)\label{eq:channelprob_3}\,,
    \end{align}
    where (\ref{eq:channelprob_2}) follows from the law of total probability and (\ref{eq:channelprob_3}) from the symmetry of the $t$-error correcting decoder. The same steps can also be applied to $\sum_{i \in \mathcal{M}} P_{Y^p|X^p}(\mathbf{y}_p|\mathbf{c}_i)$, thus giving the wanted result.
\end{proof}

\section{Proof of Proposition 3}\label{app:prop3}
\begin{proof}
    If $\mathcal{C} = \mathbb{F}_2^n$, the syndrome of any received signal vector is zero. Obviously, any hard-decision decoder can correct only up to $t = \lfloor\frac{d-1}{2}\rfloor$ errors, where $d$ is the minimum Hamming distance. In the given code the minimum Hamming distance is $d=1$ and hence, it can correct up to $t=0$ bit errors. The same is true for the subcode on $\mathcal{A}^p$. Hence, we can provide the parity-check matrix of such a code as $\mathbf{H}_p = (0,...,0)$.\\
    The syndrome of any received signal vector $\mathbf{Q}(\mathbf{y}_n)$ results to $\mathbf{0}$ and hence, the coset is given as $\mathcal{C}(\mathbf{0})= \mathbb{F}_2^n$. Furthermore, the conditional correct decoding probability results to 
    \begin{align}
        P_{\mathcal{D}_\mathrm{A}|E^p=\mathbf{e}} = \begin{cases}
            \delta&\text{if }\mathbf{e} = \mathbf{0}\,,\\
            0&\text{otherwise}\,,
        \end{cases}
    \end{align}
    where $\delta > 0$ is a positive constant that depends on the channel properties. Note that we explicitly exclude the case $\delta = 0$, which happens if and only if the bit error probability of the channel is zero. Then, using Th~\ref{th:t_parity}, the Neyman-Pearson quotient reduces to
    \begin{align}
        T_\mathrm{NP}(\mathbf{y}_p) = \delta P_{E^p|Y^p}(\mathbf{0}|\mathbf{y}_p) = \frac{\delta}{P_{\mathcal{D}_\mathrm{A}}}\prod_{i=1}^p \left(1 - \frac{1}{1+\exp{|\Lambda_i|}}\right).
    \end{align}
    We can further equivalently modify the Neyman-Pearson test with $\tau = 1$ as
    \begin{align}
        Z_\mathrm{NP}(\mathbf{y}_p) := \begin{cases}
                1&\text{if }\sqrt[p]{T_\mathrm{NP}(\mathbf{y}_p)} \geq \sqrt[p]{c}\,,\\
                0&\text{if }\sqrt[p]{T_\mathrm{NP}(\mathbf{y}_p)} < \sqrt[p]{c}\,.\\
            \end{cases}
    \end{align}
    Now, we can state a condition when the behavior of the test $Z_\mathrm{E}$ is approximately the same as the Neyman-Pearson test
    \begin{align}
        \sqrt[p]{T_\mathrm{NP}(\mathbf{y}_p)} \approx \sqrt[p]{\frac{\delta}{P_{\mathcal{D}_\mathrm{A}}}}\left(1- \frac{1}{p}\sum_{i=1}^p \frac{1}{1+\exp{|\Lambda_i|}}\label{eq:mutex_llr}\right)\,,
    \end{align}
    if for all $i = 1,2,...p$ the \acp{LLR} are approximately the same: $\Lambda_i \approx \Lambda$. Furthermore, we note that $1 - x$ and $x^{-1}$ are both strictly monotonously falling functions for $x > 0$, and hence, $Z_\mathrm{E}$ and $Z_\mathrm{NP}$ are approximately equivalent for any achievable $\alpha$-level.
\end{proof}

\fi

\end{document}